\documentclass[a4paper,11pt]{article}
\usepackage{pos}

\usepackage[export]{adjustbox}

\newcommand{\iu}{\mathrm{i}\hskip0.07em}
\newcommand{\eq}{\mathrm{eq}}

\title{Cutting rules on a cylinder and simplified diagrammatic approach to $CP$ violation in quantum kinetic theory}
\ShortTitle{Cutting rules on a cylinder and quantum kinetic theory}

\author*[a]{Peter Maták}
\author[a]{Tomáš Blažek}

\affiliation[a]{Comenius University in Bratislava,\\
 Mlynská dolina F1, 84248 Bratislava, Slovak Republic}

\emailAdd{peter.matak@fmph.uniba.sk}
\emailAdd{tomas.blazek@fmph.uniba.sk}

\abstract{We briefly explain a novel diagrammatic method for including thermal corrections in $CP$ asymmetric reaction rates entering the quantum Boltzmann equation. In thermal equilibrium, the asymmetries have to cancel precisely due to the $S$-matrix unitarity and $CPT$ invariance. Such cancelations are easy to track in zero-temperature leading-order calculations. However, accounting for medium effects requires modification of the on-shell part of propagators and, consequently, the $CP$ violating rates. In the literature, the correct form of the statistical factors in the asymmetry source term has been obtained employing the real-time-formalism of non-equilibrium field theory in a certain approximation. We demonstrate that the same results can be obtained by summing an infinite sequence of zero-temperature asymmetries. Those are derived from cuttings of forward diagrams drawn on a cylindrical surface, while thermal corrections come from windings of their propagators. The procedure is entirely diagrammatic, and the $CPT$ and unitarity cancelations are formulated for thermally corrected reaction rate asymmetries. The simplification achieved is the primary focus of our work. The aspect of infrared finiteness in higher-order corrections will also be discussed. The contribution is primarily based on our previous work in Refs. \cite{Blazek:2021olf, Blazek:2021zoj, Blazek:2021gmw}.}

\FullConference{
7th Symposium on Prospects in the Physics of Discrete Symmetries (DISCRETE 2020-2021)\\
29th November - 3rd December 2021\\
Bergen, Norway}

\begin{document}
\maketitle

\section{Introduction}

Thermal effects in leptogenesis have been previously studied within the real-time \cite{Giudice:2003jh, Garny:2009rv, Garny:2009qn, Garny:2010nj, Salvio:2011sf} as well as the imaginary-time \cite{Laine:2011pq, Bodeker:2017deo, Bodeker:2019rvr} formalisms of quantum field theory. In Refs. \cite{Biondini:2013xua, Biondini:2015gyw, Biondini:2016arl} a nonrelativistic effective theory was used to include the standard model higher-order corrections. Our work \cite{Blazek:2021olf, Blazek:2021zoj, Blazek:2021gmw} focuses on a newly introduced bottom-up approach, starting with the classical Boltzmann equation and using perturbative unitarity to include the quantum thermal effect and higher-order corrections.

\section{$CP$ asymmetries and unitarity constraints in vacuum}

To begin with, let us consider the Boltzmann equation describing the number density of a particle species labelled as $i_1$. In general, its evolution is determined by all kinematically allowed $i\rightarrow f$ processes available at the given perturbative order
\begin{align}\label{eq1}
\dot{n}_{i_1}+3H n_{i_1}= -\mathring{\gamma}_{fi}+\mathring{\gamma}_{if}+\ldots
\end{align}
where $i_1$ is assumed to be included in the initial state. The classical, or circled, reaction rates $\mathring{\gamma}_{fi}$ counting the average number of $i\rightarrow f$ reactions per unit volume per unit time are defined as \cite{Kolb:1979qa}
\begin{align}\label{eq2}
\mathring{\gamma}_{fi}=\int\prod_{i}[d\mathbf{p}_i]\mathring{f}_i(p_i)\int\prod_{f}[d\mathbf{p}_f](2\pi)^4\delta^{(4)}(p_f-p_i)\vert\mathring{M}_{fi}\vert^2
\end{align}
where $[d\mathbf{p}] = d^3\mathbf{p}/((2\pi)^3 2E_{\mathbf{p}})$. Assuming all initial-state particles are in kinetic equilibrium, the circled phase-space densities $\mathring{f}_i(p_i)$ correspond to the Maxwell-Boltzmann exponentials. The scattering amplitude in Eq. \eqref{eq2} enters the standard definition of the $S$-matrix as $S_{fi}=\delta_{fi}+\iu T_{fi}$ with $T_{fi}=(2\pi)^4 \delta^{(4)}(p_f-p_i)\mathring{M}_{fi}$ and the circle indicating the use of zero-temperature Feynman rules. 

How can we obtain a complete set of reactions affecting the $i_1$ density when considering all contributions of a specific finite perturbative order? This question may become nontrivial when higher orders are to be included. The answer follows from the diagrammatic structure of the theory  and the unitarity condition for the $S$-matrix written as
\begin{align}\label{eq4}
\iu T^\dagger_{fi} = \iu T^{\vphantom{\dagger}}_{fi} - \sum_n \iu T^{\vphantom{\dagger}}_{fn} \iu T^\dagger_{ni}.
\end{align}
Applying Eq. \eqref{eq4} iteratively, we can express the amplitude squared as \cite{Blazek:2021olf}
\begin{align}\label{eq5}
\vert\mathring{M}_{fi}\vert^2 = -\frac{1}{V_4}\iu T^{\vphantom{\dagger}}_{if} \iu T^\dagger_{fi} = 
-\frac{1}{V_4}\bigg(\iu T^{\vphantom{\dagger}}_{if} \iu T^{\vphantom{\dagger}}_{fi}
-\sum_n \iu T^{\vphantom{\dagger}}_{if} \iu T^{\vphantom{\dagger}}_{fn} \iu T^{\vphantom{\dagger}}_{\vphantom{f}ni}
+\sum_{n,m} \iu T^{\vphantom{\dagger}}_{if} \iu T^{\vphantom{\dagger}}_{fm} 
\iu T^{\vphantom{\dagger}}_{\vphantom{f}mn} \iu T^{\vphantom{\dagger}}_{ni}-\ldots
\bigg)
\end{align}
where $V_4=(2\pi)^4\delta^{(4)}(0)$ denotes the four-dimensional volume. Thus we should start with all possible vacuum diagrams made of selected vertices \cite{Blazek:2021olf, Botella:2004ks}. Cutting the $i_1$ and other internal lines will generate forward diagrams corresponding to all initial states containing $i_1$ particles. Further cutting of these diagrams in all kinematically allowed ways any number of times generates the terms of the truncated unitarity expansion in Eq. \eqref{eq5}.

Unitarity and $CPT$ constraints for $CP$ asymmetries are essential to the Sakharov conditions stating that no asymmetry can be generated in thermal equilibrium \cite{Sakharov:1967dj}. In a $CPT$ symmetric quantum theory, the asymmetries are defined as $\Delta\vert T_{fi}\vert^2 = \vert T_{fi}\vert^2 - \vert T_{if}\vert^2$ leading to \cite{Blazek:2021olf}
\begin{align}\label{eq6}
\Delta \vert T^{\vphantom{\dagger}}_{fi}\vert^2 =& \sum_{n}(\iu T_{in} \iu T_{nf} \iu T_{fi} - \iu T_{if} \iu T_{fn} \iu T_{ni})\\
&-\sum_{n,m}(\iu T_{in} \iu T_{nm} \iu T_{mf} \iu T_{fi} - \iu T_{if} \iu T_{fm} \iu T_{mn} \iu T_{ni})
+\vphantom{\sum_{n}}\ldots\,.\nonumber
\end{align}
After the summation over the final states, we can easily see $\sum_f \Delta\vert T_{fi}\vert^2 = 0$ \cite{Kolb:1979qa, Dolgov:1979mz}. In thermal equilibrium, analogous constraints for thermally averaged reaction rate asymmetries are obtained. The expansion of Eq. \eqref{eq6} is particularly useful to make the asymmetry cancelations explicit at a diagrammatic level \cite{Blazek:2021olf}, which may be understood as a higher-order generalization of the cyclic representation of Ref. \cite{Roulet:1997xa}.

\section{Perturbative unitarity at finite temperature}

Let us start with the following observation to understand the role of the $S$-matrix unitarity in the inclusion of quantum thermal corrections to the reaction rates and their $CP$ asymmetries.  Considering the seesaw type-I Lagrangian density
\begin{align}\label{eq7}
\mathcal{L}\supset -\frac{1}{2}M_i\bar{N}_iN_i-\left(Y_{\alpha i}\bar{N}_i P_L l_{\alpha}H + \mathrm{H.c.}\right),
\end{align}
how do the interactions of the standard model leptons $l$ and the Higgs doublet $H$ affect the number density of heavy right-handed neutrinos? At the leading $\mathcal{O}(Y^2)$ order, the $N_i\leftrightarrow lH, \bar{l}\bar{H}$ decays and inverse decays contribute. With Eq. \eqref{eq5} in mind, such contributions can be represented by cutting the $N_i$ self-energy diagram as
\begin{align}\label{eq8}
\includegraphics[scale=1,valign=c]{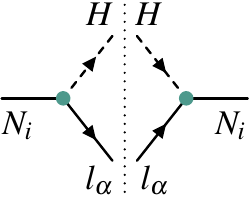}
\end{align}
and similarly for the $CP$ conjugated rate. We emphasize that it is not the only contribution at this perturbative order with the $N_i$ in the initial state. We may, for example, consider one or more Higgs particles in the initial state, leading to
\begin{align}\label{eq9}
\includegraphics[scale=1,valign=c]{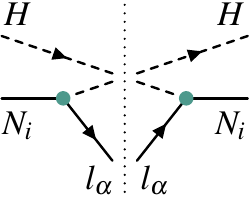}\hskip1mm,\hskip2mm\includegraphics[scale=1,valign=c]{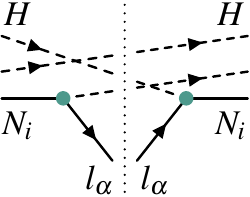}\hskip1mm,\hskip2mm\ldots
\end{align}
coming with the same amplitude squared as in Eq. \eqref{eq8}. However,  according to the definition in Eq. \eqref{eq2}, their contribution to the right-handed neutrino reaction rates includes extra powers of the circled phase-space densities. In equilibrium, the circled densities correspond to Maxwell-Boltzmann exponentials. Adding the whole sequence in Eq. \eqref{eq9} to Eq. \eqref{eq8} then leads to
\begin{align}\label{eq10}
\sum^\infty_{w=0} \left(\mathring{f}^\eq_H\right)^w = \frac{\exp\left\{E_H/T\right\}}{\exp\left\{E_H/T\right\}-1} = 1+f^\eq_H.
\end{align}
In this diagrammatic way, the Bose-Einstein quantum statistical factor has been obtained. The diagrams in Eq. \eqref{eq9} may come from drawing the simple diagram in Eq. \eqref{eq8} on a cylindrical surface, winding the Higgs line any number of times. As shown in our previous work \cite{Blazek:2021zoj}, the winding number of a bosonic particle equals to the occupation number of a single-particle state in the Fock space if the temporal evolution of a free slowly-varying density matrix is considered. Within such approximation, at any perturbative order, all quantum thermal effects are automatically included by the expansion of Eq. \eqref{eq5} considering forward diagrams of all possible winding numbers. 

In particular, thermal-corrected lepton-number asymmetry source-term derived using this procedure reads \cite{Blazek:2021zoj}
\begin{align}\label{eq11}
\int[d\mathbf{p}_{N_i}]\delta f^{\vphantom{\eq}}_{N_i} \int[d\mathbf{p}_{l}][d\mathbf{p}_{H}](2\pi)^4\delta^{(4)}(p_{N_i}-p_l-p_H)
\Big(1-f^\eq_{l}\Big)\Big(1+f^\eq_{H}\Big)\\
\int[d\mathbf{p}_{\bar{l}}][d\mathbf{p}_{\bar{H}}](2\pi)^4\delta^{(4)}(p_{l\vphantom{\bar{l}}}+p_{H\vphantom{\bar{H}}}-p_{\bar{l}}-p_{\bar{H}})
\Big(1-f^\eq_{\bar{l}}+f^\eq_{\bar{H}}\Big)\Delta \vert\mathring M\vert^2\nonumber
\end{align}
where particle momenta and phase-space distributions are labeled by subscripts, while the difference between the actual and equilibrium $N_i$ density is denoted $\delta f^{\vphantom{\eq}}_{N_i}$. Furthermore, $\Delta \vert\mathring M\vert^2$ is the zero-temperature asymmetry of the $N_i\rightarrow lH$ decay as derived in Ref. \cite{Covi:1996wh}. Remarkably, this result agrees with what has been obtained using the closed-time-path formalism of the nonequilibrium field theory \cite{Garny:2009rv, Garny:2009qn, Garny:2010nj}. The naive use of the finite-temperature generalization of the Cutkosky rules \cite{Kobes:1985kc, Kobes:1986za} would lead to incorrect statistical factors in Eq. \eqref{eq11}.

Another nontrivial example of using the expansion in Eq. \eqref{eq5} in finite-temperature calculations is the diagrammatic representation of the thermal mass effects \cite{Blazek:2021gmw}.  Including the top Yukawa interactions, one may, for example, consider a box diagram cut as
\begin{align}\label{eq12}
\includegraphics[scale=1, valign=c]{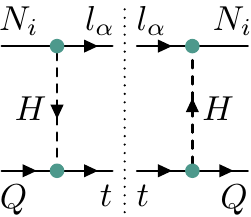}
\end{align}
or \cite{Blazek:2021gmw}
\begin{align}\label{eq13}
\includegraphics[scale=1, valign=c]{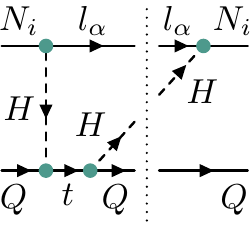}
\hskip1.5mm+\hskip1.5mm\includegraphics[scale=1, valign=c]{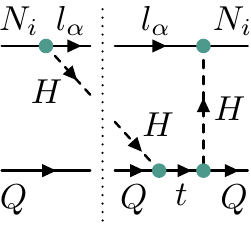}
\hskip1.5mm-\hskip1.5mm\includegraphics[scale=1, valign=c]{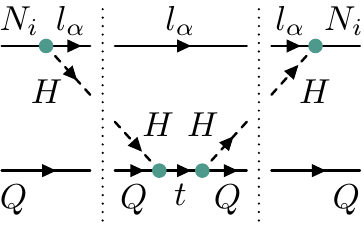}
\end{align}
representing the $N_iQ\rightarrow lt$ \cite{Nardi:2007jp,Pilaftsis:2003gt, Pilaftsis:2005rv} and $N_iQ\rightarrow lHQ$ \cite{Racker:2018tzw} reaction rates, respectively. As the quarks, leptons, and Higgses are considered massless, these rates are infrared divergent. However, their sum is finite by the Kinoshita-Lee-Nauenberg theorem \cite{Racker:2018tzw, Kinoshita:1962ur, Lee:1964is, Frye:2018xjj}. Including the complete list of contributions analogous to Eq. \eqref{eq13}, we may observe that the sum of all $2\rightarrow 3$ rates is finite as well and can be interpreted as a thermal mass effect in the $N_i\rightarrow lH$ decay \cite{Blazek:2021gmw}.

\section{Conclusions}

A newly introduced method for including quantum thermal corrections to the classical Boltzmann equation, based on the direct use of perturbative unitarity, has been briefly discussed. At the diagrammatic level, thermal-corrected reaction rates can be represented by diagrams with internal lines wound on a cylindrical surface and cut into as many pieces as possible.

\section*{Acknowledgements}
The authors were supported by the Slovak Ministry of Education Contract No. 0243/2021.

\bibliographystyle{JHEP}
\bibliography{peter_matak.bib}

\end{document}